\documentclass[twocolumn,showpacs,preprintnumbers,amsmath,amssymb]{revtex4}
\usepackage{amsfonts, amsmath, bm, bbm, graphicx, epsfig, epstopdf}
\usepackage{dcolumn}
\usepackage{textcomp}
\begin{document}


\title{Raman quantum memory based on an ensemble of nitrogen-vacancy centers coupled to a microcavity}

\author{Khabat Heshami$^{1}$, Charles Santori$^{2}$, Behzad Khanaliloo$^{1}$, Chris Healey$^{1}$, Victor M. Acosta$^{2}$, Paul E. Barclay$^{1,3}$ and Christoph Simon$^{1}$}
\affiliation{$^{1}$Institute for Quantum Science and Technology
and Department of Physics and Astronomy, University of
Calgary, Calgary T2N 1N4, Alberta, Canada\\ $^{2}$ Hewlett Packard Laboratories, 1501 Page Mill Rd., Palo Alto, CA, 94304, USA\\
$^3$ NRC National Institute for Nanotechnology, 11421 Saskatchewan Drive NW, Edmonton T6G 2M9, Alberta, Canada}

\date{\today}

\begin{abstract}
We propose a scheme to realize optical quantum memories in an ensemble of nitrogen-vacancy centers in diamond that are coupled to a micro-cavity. The scheme is based on off-resonant Raman coupling, which allows one to circumvent optical inhomogeneous broadening and store optical photons in the electronic spin coherence. This approach promises a storage time of order one second and a time-bandwidth product of order 10$^7$. We include all possible optical transitions in a 9-level configuration, numerically evaluate the efficiencies and discuss the requirements for achieving high efficiency and fidelity.
\end{abstract}

\maketitle
Quantum memories for optical photons \cite{Lvovskyrev,Tittelrev,Simonrev} are essential elements for photonic quantum information processing. Long-distance quantum communication based on quantum repeaters \cite{DLCZ,Sangouardrmp} requires optical quantum memories. They can also be used in conjunction with probabilistic photon pair sources in order to realize deterministic single-photon sources \cite{PanPRL06}, which are necessary for linear optical quantum computation \cite{KLM}. The currently leading physical realizations of quantum memories include vapor cells, rare-earth ion doped crystals and cold atoms \cite{HosseiniNatComm11,ReimNatPhot10,HedgesNat10,Sabooni13,PanNatPhys12}, where the use of ensembles \cite{Hammererrev} facilitates efficient coupling to photons. However, because of the physical dimensions or complexity of the trapping mechanism, these implementations do not seem to be well suited for miniaturization, which will be required for integrated on-chip photonic quantum information processing architectures \cite{O'Brien}.

 Negatively charged nitrogen-vacancy (NV$^-$) centers in diamond are attractive systems for implementing micron-scale optical quantum memories. NV$^-$ centers demonstrate strong coupling to optical photons, which can be further enhanced via optical microcavities \cite{LoncarFaraon}. Entanglement between an optical photon and the electronic spin of a single NV$^-$ center \cite{Togan}, and between electronic spins of two distant NV$^-$ centers \cite{Bernien} have recently been demonstrated. Ground state electronic spin coherence times of 0.6 s have been shown using dynamical decoupling \cite{T2time}. The electronic spin coherence in NV$^-$ ensembles has been used for storage and retrieval of microwave photons, see \cite{KuboPRA12}. However, storing quantum states of optical photons remains challenging.
In contrast to rare-earth ion doped crystals \cite{HedgesNat10,SaglamurekNat}, NV$^-$ centers exhibit a relatively short excited state lifetime, which prevents long-lived storage based on optical coherences.  Electromagnetically induced transparency (EIT), which is based on application of a resonant control beam in a $\Lambda$-level configuration, has been implemented \cite{Hemmer01,Acosta2013}. This could be an approach to use the ground state spin coherence to store optical photons. However, optical inhomogeneous broadening and interference due to closely-spaced excited states make it difficult to achieve high EIT contrasts \cite{Acosta2013}, and the resulting loss prevents quantum storage of optical photons.

\begin{figure}[t]
\scalebox{0.50}{\includegraphics[viewport=50 0 350 420]{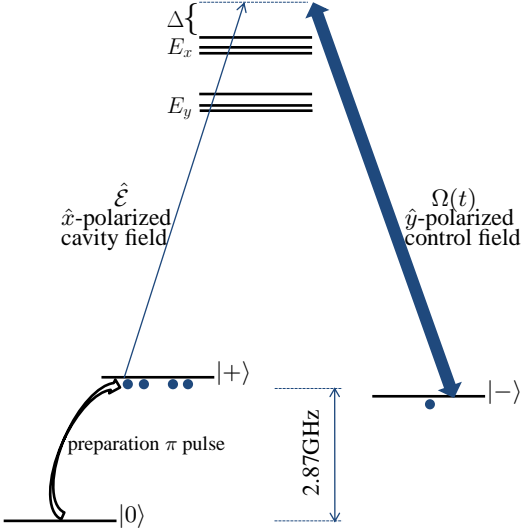}}
\caption{(Color online) Raman quantum memory scheme. The figure shows the 9-level configuration of the electronic ground and excited states for a NV$^-$ center. Initially, a microwave $\pi$-pulse transfers the population from $|0\rangle$ to $|+\rangle$. An ${\hat x}$-polarized signal photon is coupled into a cavity to be stored. For storage, a collective spin excitation is generated through off-resonant coupling to the ${\hat x}$-polarized cavity field and the ${\hat y}$-polarized control field. For retrieval, a similar control field pulse is applied to read out the stored excitation. $\Delta$ is the cavity and control field detuning from the excited state.} \label{figure1}
\end{figure}

In this letter, we propose to use an off-resonant Raman coupling approach \cite{NunnPRA07,ReimNatPhot10}  that allows one to circumvent the excited state inhomogeneous broadening, see Fig. \ref{figure1}. In our scheme, we consider an ensemble of NV$^-$ centers coupled to an optical microcavity. The NV$^-$ ensemble is initialized in a ground state and interacts with a cavity field and a control field pulse. For storage, the input field is coupled to the cavity and the control field pulse is simultaneously applied to the ensemble. This results in storing the optical photon and generating a collective spin excitation. For retrieval, one can apply a similar control field pulse to read out the spin excitation and generate a photon in the cavity's output. The NV$^-$ ensemble is considered under a very high external static electric field and a low magnetic field in order to achieve the desired optical polarization selection rules.

We now describe the level structure of NV$^-$ centers. Each NV$^{-}$ center consists of six contributing electrons in the $C_{3v}$ symmetry that is imposed by the diamond crystal lattice \cite{DohertyNJP,Maze}. The electronic configuration consists of six excited states and one ground state triplet. Neglecting hyperfine coupling with the nuclear spin, the ground eigenstates under external electric and magnetic fields can be shown to be
\begin{eqnarray}\label{gstates}
&& |0\rangle=|S=1,m_s=0\rangle,\\ \nonumber
&& |+\rangle=e^{i\phi_e/2}\cos\left(\frac{\theta}{2}\right)|1,1\rangle -e^{-i\phi_e/2}\sin\left(\frac{\theta}{2}\right)|1,-1\rangle,\\ \nonumber
&& |-\rangle=e^{i\phi_e/2}\sin\left(\frac{\theta}{2}\right)|1,1\rangle +e^{-i\phi_e/2}\cos\left(\frac{\theta}{2}\right)|1,-1\rangle,
\end{eqnarray}
where $\phi_e=\arctan\left(\frac{E_y^{gs}}{E_x^{gs}}\right)$, $\theta=\arctan\left(\frac{E_{\perp}^{gs}}{B_z^{gs}}\right)$ and $E_{\perp}^{gs}=\sqrt{{E_x^{gs}}^2+{E_y^{gs}}^2}$, see \cite{DohertyPRB}. $E_{x,y}^{gs}$ and $B_z^{gs}$ are the ground-state energy shifts due to electric and magnetic fields according to the relevant dipole moments and g-factors \cite{Doherty13}. The energy splitting between $|\pm\rangle$ states is given by $2\sqrt{{E_{\perp}^{gs}}^2+{B_z^{gs}}^2}$. In the above description $|S=1,m_s=0,\pm 1\rangle$ are the NV$^{-}$ center's ground configuration states $\Phi_{A_2,1,\{0,\pm 1\}}^c$, where $S=1$ shows the total spin for the NV$^-$ center's ground state triplet, see Table 1 in \cite{DohertyNJP}. Likewise, one needs to consider the effect of external static electric and magnetic fields on the excited states. For this purpose, we derive spin-orbit and spin-spin interaction Hamiltonians in the configuration basis for the excited state triplets, see the supplementary information \cite{supplementary}.

Originally, an ensemble of oriented NV$^{-}$ centers, see below, are prepared in the $\Phi^{c}_{A_2;S=1,m_s=0}$ ground state using off-resonant optical pumping \cite{pumping}. As shown in Fig. \ref{figure1}, a preparation microwave $\pi$-pulse prepares all NV$^-$ centers in the $|+\rangle$ (with $m_I=0$)  ground state. An imperfect preparation may result in either exciting some of the NV$^-$ centers to the $|-\rangle$ state or addressing multiple hyperfine levels corresponding to the $|+\rangle$ state. Since our regime of parameters will result in electronic spin splittings larger than the hyperfine splitting, the bandwidth of the preparation $\pi$-pulse should be narrow compared with the hyperfine splitting of 2.2 MHz to avoid coupling to multiple hyperfine levels. At the same time, the microwave pulse bandwidth should at least be comparable with the spin inhomogeneous broadening of ~200 kHz (see below).

\begin{table}[t]
\caption{The following table shows coupling ratios $|g_{x}(j,k)/g_{x}(+,9)|$ for coupling to an ${\hat x}$-polarized light. The electric and magnetic field splittings are, $E_{x}^{es}=120GHz, B_{z}^{es}=10kHz$ and $E_{y,z}^{es}=B_{x,y}^{es}=0$.\label{table1}}
\begin{tabular}{| l || c | c | c | c | c | c|}
    \hline
            & k=4 & k=5 & k=6 & k=7 & k=8 & k=9 \\ \hline
    $j=1,|0\rangle$& $<10^{-4}$ & $<10^{-4}$ & 0.0006 & 1.0003 & $<10^{-4}$ & 0.0585 \\ \hline
    $j=2,|+\rangle$& $<10^{-4}$ & $<10^{-4}$ & 0.0253 & 0.0585 & 0.0015 & 1 \\ \hline
    $j=3,|-\rangle$& 0.0050 & 0.0182 & $<10^{-4}$ & 0.0001 & 1.0019 & 0.0015\\
    \hline
  \end{tabular}
\end{table}

It is crucial to determine polarization selection rules in order to study all active transitions in the light-NV$^{-}$ interaction Hamiltonian. We consider ground eigenstates in Eq. (\ref{gstates}) and excited states that are derived from Eqs. (S1,S2) in the supplementary information \cite{supplementary}. Taking into account that $\langle a_1|{\hat x}\cdot{\vec r}|e_x\rangle$ and $\langle a_1|{\hat y}\cdot{\vec r}|e_y\rangle$ are non-zero \cite{Maze}, where $a_{1,2}$ and $e_{x,y}$ denote single electron orbital basis given by linear combination of the dangling orbitals, we find all possible optical transitions for ${\hat x},{\hat y}$ polarized light from ground eigenstates to any of the excited states, see Table \ref{table1} and \cite{supplementary}. At a very low magnetic field, an external electric field mixes $\Phi_{A_2;1,\pm 1}^{c}$ ground states to $|\pm\rangle$ states that are shown in Eqs. (\ref{gstates}). In addition, the external electric field (in the $x$ direction, where the $z$ direction is defined by orientation of the NV$^-$ centers and $x$ is along one of the reflection planes) results in splitting of the excited states to the $E_x$ and $E_y$ branches, \cite{Maze,DohertyNJP,NVStark}. As a result, one can couple transitions from $|\pm\rangle$ states to the excited states through linearly polarized photons. This is crucial to avoid complications in coupling NV$^{-}$ centers to laterally confined cavities such as micro-ring and photonic crystal cavities that generally lack polarization degeneracy in their modes (in contrast with planar Fabry-Perot-type cavities which can have polarization degeneracy).

Let us describe the dynamics of this system. First, free evolution of the system is given by $H_0=\hbar\omega_c a^{\dag}a + \sum_{i=1}^{N}\sum_{j=1..9} e_j^i {\hat \sigma}_{jj}^{i},$
where $\omega_c$ is the cavity's central frequency, $a$($a^{\dag}$) is the cavity photon's annihilation(creation) operator and ${\hat \sigma}_{\mu\nu}^i=|\mu\rangle^i\langle\nu|$. The eigenenergies of the $i$th NV$^-$ center $e_j^i$ are given by ground and excited state Hamiltonians, where $j=1\ldots 3$ refers to $|0\rangle$, $|+\rangle$ and $|-\rangle$ ground states, respectively. $j=4\ldots 9$ denote the excited eigenstates, where $j=4$ and $j=9$ refer to the lowest and highest energy excited states. In the present scheme, NV$^-$ centers interact with $x$-polarized cavity and $y$-polarized control fields. The interaction Hamiltonian is given by,
\begin{eqnarray}\label{Hint}
&&H_{int}=-\hbar \sum_{i=1}^{N}\sum_{j=1..3}\sum_{k=4..9} {\hat {\cal E}}G(j,k){\hat \sigma}_{kj}^{i} e^{-i\omega_c t} \\ \nonumber
&& + \Omega(j,k){\hat \sigma}_{kj}^{i} e^{-i\omega_2 t}+ h.c.,
\end{eqnarray}
where ${\hat {\cal E}}=ae^{i\omega_c t}$, $G(j,k)=g_x(j,k) d_{zpl}\sqrt{\frac{\omega_c}{2\hbar \epsilon V}}$ and $\Omega(j,k)= d_{zpl}g_y(j,k)\frac{E_2}{2\hbar}$. $E_2$ and $\omega_2$ are amplitude and frequency of the control field, and
$g_{x,y}(j,k)=\frac{\vec{\mu}_{jk}\cdot\hat{x},\hat{y}}{|\mu_{jk}|}$, where $\vec{\mu}_{jk} = \langle j|\vec{r}|k\rangle$.
The transition dipole moment of the zero phonon line (zpl) is given by $d_{zpl}=\sqrt{\frac{3\pi^2\epsilon_0 \hbar c^3 \gamma_{zpl}}{n_d \omega_0^3}}$, where $n_d$ is diamond's refractive index and $\omega_0$ is the transition frequency that is associated with $\lambda=637$ nm. The above definition of $d_{zpl}$ is based on $\gamma_{zpl}=0.035\gamma$, where $\gamma$ is the radiative decay rate \cite{Doherty13}. This takes into account that only few percent of the emission from the excited state is associated with the zero-phonon line. Note that the broad phonon sidebands do not affect the proposed process as they are a few nanometers detuned from the inhomogeneously broadened zero phonon line.

One can rewrite the total Hamiltonian in terms of collective optical polarization operators ${\hat \sigma}_{(1,2)k}=\sum_{i=1}^{N} {\hat \sigma}_{(1,2)k}^i e^{i\omega_c t}$, ${\hat \sigma}_{3k}=\sum_{i=1}^{N} {\hat \sigma}_{3k}^i e^{i\omega_2 t}$ and ${\hat \sigma}_{\mu\mu}=\sum_{i=1}^{N} {\hat \sigma}_{\mu\mu}^i$, where $N$ is the total number of NV$^-$ centers, $k=4\ldots 9$ and $\mu=1\ldots 9$. Using the Heisenberg equation, ${\dot {\hat {\cal O}}}=\frac{i}{\hbar} [H, {\hat {\cal O}}]+\frac{\partial {\hat {\cal O}}}{\partial t}$, we find the dynamics of the cavity field operator and the collective operators describing the spin and optical polarizations and populations, see the supplementary information \cite{supplementary}.
In our model, we include relevant decay and decoherence rates. Specifically, optical inhomogeneous broadening ($\gamma_{e}$), spin inhomogeneous broadening ($\gamma_s$) and excited state radiative decay ($\gamma$) are included in the dynamics of the optical polarizations, spin polarization and level populations.
Nonlinear contributions in these dynamical equations for collective operators can be ignored if the number of NV$^-$ centers is much larger than the number of input photons \cite{Gorshkov,HeshamiPRA11}. Due to linearity of the dynamics, one can solve the same dynamical equations to find solutions to the single excitation wavefunctions of the corresponding collective operators \cite{Hammererrev,Gorshkov}, see below. For example, the single spin excitation wavefunction is given by the dynamics of ${\hat \sigma}_{23}$.

One can also derive the dynamics of the cavity field operator. Given that the cavity decay rate is the fastest rate in the system, we can use adiabatic elimination to simplify the cavity field dynamics \cite{HeshamiPRA11}. This leads to
\begin{eqnarray}\label{Ecav}
&&{\hat {\cal E}}(t)=\frac{1}{\kappa}\{ \sqrt{2\kappa} {\hat {\cal E}}_{in}(t)+i\sum_{j=1,2}\sum_{k=4..9}G^*(j,k){\hat \sigma}_{jk} \\ \nonumber
&&+ i\sum_{k=4..9}G^*(3,k){\hat \sigma}_{3k} e^{-i(\omega_2-\omega_c)t}\},
\end{eqnarray}
where ${\hat {\cal E}}_{in}(t)$ is the annihilation operator corresponding to the input signal. The cavity input-output equation, ${\hat {\cal E}}_{out}(t)=-{\hat {\cal E}}_{in}(t)+ \sqrt{2\kappa}{\hat {\cal E}}(t)$, in combination with the above considerations allows one to analyze the proposed memory scheme and study its performance, see \cite{Gorshkov,HeshamiPRA11} for similar treatments.

\begin{figure}[t]
\epsfig{file=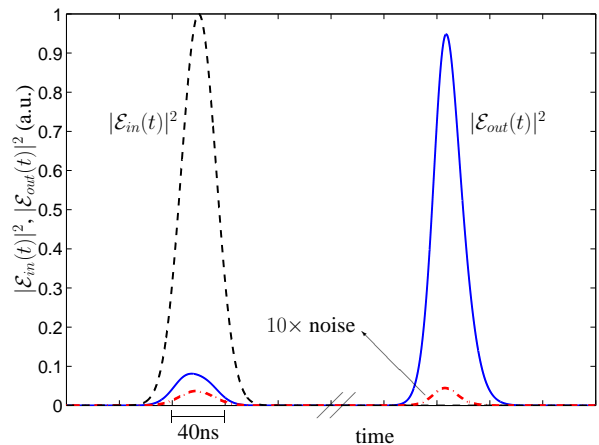,width=0.9\columnwidth}
\caption{(Color online) Simulation results of storage and retrieval of input pulse (black dashed line). The blue solid line presents the cavity output field.
Simulation parameters: pulse bandwidth $\Delta\omega$=110 MHz, excited state inhomogeneous broadening, $\gamma_e$=1 GHz, spin inhomogeneous broadening $\gamma_s$=200 kHz. The detuning from the excited state, $\Delta=0.8$ GHz (other parameters are provided in the main text). This results in 91\% absorption efficiency and ~81\% total efficiency. The red dash-dotted line is associated with the noise intensity (multiplied by 10 for clarity), corresponding to 1\% total noise probability. This gives 99\% conditional fidelity, see text for more details.} \label{inout}
\end{figure}

The total efficiency is found based on $\eta_{tot}=\frac{\int|{\cal E}_{out}(t)|^2dt}{\int|{\cal E}_{in}(t)|^2dt}$, where ${\cal E}_{in,out}(t)=\langle 0|{\hat {\cal E}}_{in,out}(t) |1 \rangle$ is the single photon wave function. Here $|1\rangle$ is the single photon input state. One can also find the storage efficiency by using  $\eta_{s}=1-\frac{\int dt |{\cal E}^s_{out}(t)|^2}{\int dt |{\cal E}_{in}(t)|^2}$, where ${\cal E}^s_{out}(t)$ is the field that is lost during storage. This is justified as there will be almost no population in the excited states during the storage process and therefore the radiative decay does not introduce a significant loss channel.

We consider generated noise at the output that is due to non-zero coupling of the control field to $|+\rangle\rightarrow|k\rangle$ transitions, where $|k\rangle$ refers to any of the excited states. Note that contributions from the lower branch of the excited states are suppressed due to a significant splitting of 240 GHz (see below). We calculate the noise by finding the output field in the absence of input field, which gives rise to a non-zero probability of detecting a photon at the output. This results in reduction of the storage fidelity. Using total probabilities for reading out the signal and the noise we can estimate the conditional fidelity based on  $1-\frac{P_{noise}}{P_{sig}}$, see \cite{Simonrev}.

In Fig. \ref{inout}, we show the results of our numerical solution to the differential equations for single collective excitation wavefunctions and Eq. (\ref{Ecav}) based on the assumption that all of the NV$^{-}$ centers are initially in the $|+\rangle$ state. We show a Gaussian input signal intensity, $|{\cal E}_{in}(t)|^2$, which has pulse duration of $\tau=$40 ns at $1/e$ of the maximum of the intensity. We find that this 40 ns input pulse is being stored with a storage efficiency of $\eta_{s}\approx 91\%$. For retrieval, the state of the stored spin excitation is used as the initial state of the dynamics with no input field present. Using the control field, $\Omega(t)$, we can read the stored excitation out. For the above mentioned parameters we calculate the total efficiency of $\eta_{tot}=81\%$. The total efficiency can be enhanced by increasing the control field strength. However, in the regime where $\frac{\Omega^2}{\Delta}\tau\approx 1$ the AC Stark shift affects the output field shape. The AC Stark shift can be compensated by a proper phase modulation on the control pulse \cite{ReimNatPhot10,Surmacz}. This allows one to approach the ideal retrieval efficiency \cite{Gorshkov}. In practice, a non-ideal fiber-to-cavity coupling could result in another source inefficiency. Recently, a highly efficient optical fiber to cavity coupling has been shown, see \cite{Painter13}.

Our results in Fig. \ref{inout} depend on several physical parameters including the external electric and magnetic fields, which determine the transition dipole moments, in addition to the characteristics of the ensemble and the cavity. The performance of the quantum memory scheme also depends on the detuning, $\Delta$, because of the coupling to neighboring excited states. The large number of parameters and complications due to the level structure make a systematic optimization very difficult. Below, we explain physical requirements for achieving the shown results.


Coupling of the control field to the $|+\rangle\rightarrow|k=8\rangle$ transition, see Table I in \cite{supplementary}, may result in noise through off-resonant scattering of a ${\hat x}$-polarized photon via $|k=8\rangle\rightarrow|-\rangle$ transition. In order to suppress this effect, $\Delta$ cannot be much larger than the energy splitting of these two excited states ($j=8,9$) that is about 1.5 GHz. Here, we assume $\Delta=0.8$ GHz (as shown in Fig. \ref{figure1}) and an optical inhomogeneous broadening of 1 GHz. In \cite{Santori06}, authors present a sample with NV$^-$ density of about 8000 NV/($\mu$m)$^3$ that have an optical linewidth at FWHM of 10 GHz. Here, we require a minimum NV ensemble density of about 50 NV/($\mu$m)$^3$ oriented NV$^-$ centers (this is based on number of NV$^-$ centers that are assumed for the simulation that are given below). One can employ spectral hole burning techniques \cite{hburning} to reduce the optical inhomogeneous broadening down to the 1 GHz range. In \cite{orientedNV}, production of ensemble of preferentially oriented NV$^-$ centers has been shown. Further development of this work will allow to suppress one of the two preferential orientations, see \cite{orientedNV}. In addition, in the high electric field (strain) regime that we are interested in, an appropriately chosen electric field direction may allow one to suppress contributions from the unwanted orientations of the NV$^-$ centers.

For the numerical simulation in Fig. \ref{inout}, we set $N=100$, and for this size of an ensemble, only a relatively moderate cavity quality factor of $Q=1100$ is required. The correspondingly modest density of NV$^-$ centers is compatible with achieving low optical and spin inhomogeneous broadening. Our choice of the cavity quality factor is well-justified as higher cavity quality factors have been achieved for a single NV$^-$ center in a cavity; see \cite{Faraon11,Faraon13}. The assumed cavity quality factor results in a cavity amplitude decay rate of $\kappa/2\pi=\frac{\omega_c}{2Q}=210$ GHz, which is by far the fastest rate in the system and justifies the adiabatic elimination in the derivation of Eq. (\ref{Ecav}). The mode volume is $V=100(\frac{\lambda}{n_d})^3$. A maximum Rabi frequency of $(\Omega_0/2\pi)^2=1$ (GHz)$^2$ per 1mW of control field power can be achieved for a beam waist of 8 $\mu$m. Here we considered maximum control field powers of 0.8 mW and 6.7 mW for read-in and read-out, respectively, and the control field pulse shape for storage and retrieval are set to be identical to the signal.
The cavity must efficiently couple to ${\hat x}$-polarized light (polarization of input/output signals). The ${\hat y}$-polarized control field is simultaneously applied from a different direction. Having opposite polarizations for the signal and control fields is beneficial for the scheme as it prevents excessive noise that depends on the power of the control field and lowers the conditional fidelity.

The polarization selection rules are determined by the external electric and magnetic fields. Here, $E_y^{es}=E_y^{gs}=0$ and $E_x^{es}=$120 GHz, with corresponding $E_x^{gs}=3.4$ MHz. A low magnetic field strength is assumed such that it causes $B_z^{es}=10$ kHz and $B_z^{gs}=9.9$ kHz splittings. These give the splitting of approximately 6.8 MHz between $|+\rangle$ and $|-\rangle$ ground states, see Eqs. (\ref{gstates}). Energy shifts of 17 Hz/(V/cm) and 2.8 MHz/G are expected for non-axial electric field and axial (parallel to the NV axis) magnetic field. The non-zero magnetic field is advantageous because of creating an imbalance between couplings to the two highest-energy (competing) excited states, see Table 1 and 2 in the supplementary information \cite{supplementary}, which is in favor of storage with high efficiency and fidelity. According to these coupling coefficients, a magnetic field of about 3.5 mG and an electric field of 20 V/$\mu$m will be required to achieve the above-mentioned energy shifts. It has to be noted that the energy shift due the electric field can be applied by employing a properly oriented external strain.

Here we assumed the spin inhomogeneous broadening of 200 kHz, see \cite{Acosta2013}. This relatively narrow spin inhomogeneous broadening provides a storage time of 200 ns without a significant impact on the retrieval efficiency (reduced by a factor of 0.96) without application of any rephasing $\pi$-pulse. Applying a dynamical decoupling pulse sequence, such as a series of rephasing $\pi$-pulses, can extend the storage time by many orders of magnitude \cite{LongdellPRL05,HeshamiPRA11}. The longest spin coherence lifetime measured to date using dynamical decoupling is 0.6 s \cite{T2time}. Thus, a time-bandwidth product up to $10^7$ may be possible in our scheme.

In conclusion, we proposed a scheme based on off-resonant Raman coupling for storage of optical photons in an ensemble of NV$^-$ centers that are coupled to a microcavity. High efficiencies are possible with realistic parameters, and using dynamical decoupling techniques, we expect that long storage times can simultaneously be achieved. The realization of an on-chip, efficient and long-storage-time optical quantum memory is therefore feasible owing to recent advances in NV technology. Recent results on coupling the ground-state electronic spin of NV ensembles to superconducting flux qubits in combination with the present proposal might provide a foundation for a hybrid architecture \cite{RableHybrid} that is capable of quantum communication and information processing using photons, NV$^-$ centers and superconducting circuits \cite{KuboPRL11,coherentNV-Flux,Nori}.

{\it Acknowledgments.} This work was supported by NSERC, AITF and in part by the DARPA Quiness Program under Grant No. W31P4Q-13-1-0004. K.H. would like to thank S. Raeisi and R. Ghobadi for useful discussions.

\pagebreak

\widetext
\begin{center}
\textbf{\large Supplemental Material for ``Raman quantum memory based on an ensemble of nitrogen-vacancy centers coupled to a microcavity"}
\end{center}

\setcounter{table}{0}
\setcounter{equation}{0}
\setcounter{page}{1}
\makeatletter
\renewcommand{\bibnumfmt}[1]{[SI#1]}
\renewcommand{\citenumfont}[1]{SI#1}

\section{NV$^{-}$ centers in high strain regime}
In this section we present the Hamiltonian for NV$^{-}$ centers under external static electric and magnetic fields. We use the Hamiltonian to derive optical transitions for ${\hat x}$ and ${\hat y}$ polarizations. 

Table 1 in \cite{DohertyNJP} shows the spin-orbit and configuration basis states for a negatively charged NV. Using table 2 and 3 in \cite{DohertyNJP}, we derive the spin-orbit and spin-spin interaction Hamiltonian in the configuration basis for the excited state triplets that is represented by $\{\Phi^{c}_i\}_{i=1..6}=\{ \Phi^{c}_{E,x;1,0},  \Phi^{c}_{E,x;1,1},  \Phi^{c}_{E,x;1,-1}, \Phi^{c}_{E,y;1,0}, \Phi^{c}_{E,y;1,1}, \Phi^{c}_{E,y;1,-1}\}$. 
The following Hamiltonian is the resulting spin-orbit and spin-spin Hamiltonian.
\begin{equation}\tag{S1}\label{Hsoss}
H_{so,ss}=
\begin{pmatrix}
-2D_{2A_1} & \frac{D_{2E_2}}{\sqrt{2}} & -\frac{D_{2E_2}}{\sqrt{2}} & 0 & -\frac{iD_{2E_2}}{\sqrt{2}} & -\frac{iD_{2E_2}}{\sqrt{2}} \\

\frac{D_{2E_2}}{\sqrt{2}} & D_{2A_1} & D_{2E_1} & \frac{iD_{2E_2}}{\sqrt{2}} & i\lambda_{\parallel} & -iD_{2E_1} \\

-\frac{D_{2E_2}}{\sqrt{2}} & D_{2E_1} & D_{2A_1} & \frac{iD_{2E_2}}{\sqrt{2}} & iD_{2E_1} & -i\lambda_{\parallel} \\

0 & -\frac{iD_{2E_2}}{\sqrt{2}} & -\frac{iD_{2E_2}}{\sqrt{2}} & -2D_{2A_1} & -\frac{D_{2E_2}}{\sqrt{2}} & \frac{D_{2E_2}}{\sqrt{2}} \\

\frac{iD_{2E_2}}{\sqrt{2}} & -i\lambda_{\parallel} & -iD_{2E_1} & -\frac{D_{2E_2}}{\sqrt{2}} & D_{2A_1} & -D_{2E_1}\\

\frac{iD_{2E_2}}{\sqrt{2}} & iD_{2E_1} & i\lambda_{\parallel} & \frac{D_{2E_2}}{\sqrt{2}} & -D_{2E_1} & D_{2A_1}  \\
\end{pmatrix}
\end{equation}
where $D_{2A_1}=1.42/3$GHz, $D_{2E_2}=0.2/{\sqrt{2}}$GHz, $D_{2E_1}=1.55/2$GHz and $\lambda_{\parallel}=5.3$GHz are representing the spin-spin and spin-orbit interactions. Note that there are differences between some of the coefficients in this Hamiltonian and Eq. (19) in \cite{DohertyNJP}. Despite having similar eigen-energies, differences in eigenstates can be crucial for deriving the correct optical polarization selection rules.

In this basis the Hamiltonian for external electric and magnetic fields acquires the following simple form 
\begin{equation}\tag{S2}\label{Helecmag}
H_{elec,mag} =
\begin{pmatrix}
 -E_x^{es} & 0 & 0 & E_y^{es} & 0 & 0 \\
 0 & -E_x^{es}+B_z^{es} & 0 & 0 & E_y^{es} & 0 \\
 0 & 0 & -E_x^{es}-B_z^{es} & 0 & 0 & E_y^{es} \\
 E_y^{es} & 0 & 0 & E_x^{es} & 0 & 0 \\
 0 & E_y^{es} & 0 & 0 & E_x^{es}+B_z^{es} & 0 \\
 0 & 0 & E_y^{es} & 0 & 0 & E_x^{es}-B_z^{es} \\ 
\end{pmatrix}
\end{equation}
where $B_z^{es}$ is the energy shift due to the axial magnetic field and $E_{x,y}^{es}$ are the excited-state energy shifts associated with external electric field components in the $x-y$ plane perpendicular to the NV axis . We diagonalize the total Hamiltonian ($H_{so,ss}+H_{elec,mag}$) in order to find the energies of the excited levels and their states as a superposition of the spin-orbit states (the basis).

\section{Optical transition (selection rules)}
Based on all of the above considerations and the results in Eq. (1) in the paper, we find the polarization selection rules from ground states to the excited states in the high electric field and low magnetic field condition, such that the $|S=1,m_s=\pm 1\rangle$ states are mixed, i.e. $E_{\perp}^{gs}\gg B_z^{gs}$. This approach allowed us to include all optical transitions due to the coupling to cavity and control fields in the present scheme.

For this purpose, we find excited eigenstates from the total Hamiltonian for the excited states of NV$^-$ centers that includes $H_{so,ss}$ and $H_{elec,mag}$, see Eqs. (\ref{Hsoss},\ref{Helecmag}). The $k^{th}$ excited eigenstate can be represented as $|k\rangle=\sum_{i=1..6} c_{i}^{k}\Phi^c_i$, where $\Phi^c_i$ denotes the configuration states shown above. In \cite{DohertyNJP}, these states are shown in terms of $|a_1 {\bar a}_1 e_x {\bar e}_x e_y {\bar e}_y\rangle$ states, where $\{a_1,e_x,e_y\}$ are single electron orbitals and the overbar denotes spin-down. Using this representation for the ground and excited eigenstates and  $\langle a_1|{\hat x}\cdot{\hat r}|e_x\rangle = \langle a_1|{\hat y}\cdot{\hat r}|e_y\rangle$, we examine all possible optical transitions from ground to excited states. The following table show coupling ratios $|g_{y}(j,k)/g_{y}(-,6)|$ for transitions from ground states $|0,\pm\rangle$ to excited states $|k\rangle$ for coupling to $y$-polarized light. Similar results for coupling ratios $|g_{x}(j,k)/g_{x}(+,6)|$ for $x$-polarized light are presented in Table I in the paper.

\begin{table}[h]
\caption{The following table shows coupling ratios $|g_{y}(j,k)/g_{y}(-,9)|$, where $g_{y}(j,k)=\frac{\vec{\mu}_{jk}\cdot\hat{y}}{|\mu_{jk}|}$ and $\vec{\mu}_{jk} = \langle j|\vec{r}|k\rangle$. The electric and magnetic field splittings are, $E_{x}^{es}=120GHz, B_{z}^{es}=10kHz$ and $E_{y,z}^{es}=B_{x,y}^{es}=0$.}
\begin{tabular}{| l || c | c | c | c | c | c|}
    \hline
            & $|k=4\rangle$ & $|k=5\rangle$ & $|k=6\rangle$ & $|k=7\rangle$ & $|k=8\rangle$ & $|k=9\rangle$ \\ \hline
    $j=1,|0\rangle$& 38.5440 & 9.3371 & $<10^{-4}$ & $<10^{-4}$ & 0.0229 & $<10^{-4}$ \\ \hline
    $j=2,|+\rangle$& 9.3350 & 38.5372 & 0.0584 & $<10^{-4}$ & 0.7473 & 0.0015 \\ \hline
    $j=3,|-\rangle$& 0.0137 & 0.0567 & 39.6461 & 0.0827 & 0.0011 & 1\\
    \hline
  \end{tabular}
\end{table}
The significant splitting of 240 GHz between the upper and lower branches of the excited suppresses off-resonant couplings to the lower branch due the relatively small detuning of ~0.8 GHz.
\section{Dynamical equations for optical polarizations and spin excitation}
The following shows examples from a larger set of dynamical equations that are derived from the Heisenberg equations. Here, we include the spin and optical inhomogeneous broadenings in the equations. A similar approach in \cite{Gorshkov} has been used to analyze storage in a 3-level configuration. The dynamics of the optical polarization for $|+\rangle\rightarrow |k=9\rangle$ transition is as follows
\begin{equation}\tag{S3}
{\dot {\hat \sigma}}_{29}=(i\Delta -\gamma/2-\gamma_{e})  {\hat \sigma}_{29}+ iG(2,9)N{\hat {\cal E}} +i G(3,9){\hat {\cal E}} {\hat \sigma}_{23}e^{i\delta_g t} + i\Omega(2,9)Ne^{-i\delta_g t} + i \Omega(3,9) {\hat \sigma}_{23},
\end{equation}
where $\delta_g=\omega_2-\omega_c$. This approach is adequate for treating the inhomogeneous broadening in the present context \cite{Gorshkov}. For treatment of echo rephasing pulses, inhomogeneous broadening must be modeled by explicitly including distribution of frequencies.
Similarly, the spin polarization dynamics is given by
\begin{equation}\tag{S4}
{\dot {\hat \sigma}}_{23}=-\gamma_{s}  {\hat \sigma}_{23} - i\Sigma_{k=4..9} G(2,k){\hat {\cal E}} {\hat \sigma}_{k3} + \Omega(2,k)  {\hat \sigma}_{k3}e^{-i\delta_g t} - {\hat {\cal E}}^{\dagger}G^*(3,k) {\hat \sigma}_{2k}e^{-i\delta_g t} - \Omega^*(3,k){\hat \sigma}_{2k}.
\end{equation}
For finding the single photon wavefunction of the cavity field based on the Eq. (3) in the paper, we require to find a solution to single excitation wavefunctions of spin and optical polarizations. This is performed through integrating over all dynamical equations in a this 9-level configuration and assuming that all of the NV$^-$ centers are initialized in the $|+\rangle$ ground state.

\end{document}